\documentclass[11pt]{article}

\usepackage{epsfig}
\usepackage{amssymb}
\usepackage{fullpage}

\title{An exact expression for the image error in a catadioptric sensor}
\bigskip

\author{R. Andrew Hicks\\
Department of Mathematics and Computer Science\\
Drexel University\\
ahicks@math.drexel.edu\\}

\date{}

\begin{document}
\maketitle

\abstract{A catadioptric sensor induces a projection between a given object surface and an image plane. The prescribed projection problem is the problem of finding a catadioptric sensor that realizes a given projection. Here we present a functional that describes the image error induced by a given mirror when compared with a given projection. This expression can be minimized to find solutions to the prescribed projection problem. We present an example of this approach, by finding the optimum spherical mirror that serves as a passenger side mirror on a motor vehicle.
}

\newpage

\section{Introduction}

A catadioptric sensor generally consists of a camera pointed at a
convex mirror.  Catadioptric sensors of this type tend to have large fields of view, and hence their most common application is panoramic imaging.

We will assume a simple
model of the dioptric component, known as the {\bf pinhole model},
which realizes the {\bf perspective projection}. A limiting case of perspective projection is the {\bf orthographic projection}, in which the rays that impinge upon the image plane are parallel. Such systems can be thought of as very narrow
field perspective devices.

Almost all work on catadioptric sensor design refers to sensors
employing rotationally symmetric mirrors, since these mirrors are the
simplest to make and to mathematically model. The design of such
mirrors reduces to solving an ordinary differential equation.

The earliest example of a camera employing a mirror is due to
A.S. Wolcott \cite{wolcott1840} and appears in the 1840 patent
``Method of taking Likenesses by Means of a Concave Reflector and
Plates so Prepared as that Luminous or other Rays with act Thereon.''
Remarkably, this device, which is designed for use with a
daguerreotype plate, appears just a few years after the invention of
photography. Since this patent, an enormous number of catadioptric
cameras and projection devices have appeared, many of which are
documented on the webpage \cite{hicks03designpage}, which is a
historical resource on catadioptric sensor design created by the author. The first use of differential methods for the design of a mirror
shape in a catadioptric sensor appears in the 1945 patent of Benford
\cite{benford45}. Other works that make use of differential methods
include \cite{bruckstein96}, \cite{chahl97}, \cite{baker98iccv},
\cite{hicks99cvpr}, \cite{ollis99}, \cite{conroy99iccv} ,
\cite{gaechter01icar}, \cite{hicks02omnivis},\cite{hicks03arxiv}, \cite{swaminathan03procams}. Early applications to
robotics include \cite{yagi90} and \cite{yamazawa93}. A heuristic approach to image error is discussed in \cite{swaminathan04omnivis}.

\section{Statement of the Prescribed Projection Problem} 

\label{sec:prescribed}

In this section we state the prescribed projection problem, which is
our fundamental problem of interest. Suppose one is given a fixed surface, $S$,
in $R^3$, which we will call the {\bf object surface} and a camera
with {\bf image plane} $I$, also fixed in $R^3$. A given mirrored
surface $M$ induces a transformation $T_M$ from some subset of $I$ to
$S$ by following a ray (determined by the camera model) from a point
${\bf q} \in I$ until it intersects the mirror at a point ${\bf
r}$. The ray is then reflected according to the usual law that the
angle of incidence is equal the angle of reflection and intersects $S$
at a point ${\bf s}$. We then define $T_M({\bf q}) = {\bf s}$ .

\noindent
The prescribed projection problem for systems containing a single mirror is:
\medskip
 
\noindent
\fbox{
\parbox{5.5in}{Given $G:I \longrightarrow S$, find $M$ such that
$T_M=G$. If no such $M$ exists, then find $M$ such that $T_M$ is a good
approximation to $G$.}  }
\medskip

\noindent
We will refer to $G({\bf q})$ as the {\bf target point}. If an exact
solution to the problem exists, then there are several ways to
calculate it. Otherwise, there are numerous ways to formulate and
solve the approximation problem.

Notice that for a
given $M$, with ${\bf q, r}$ and ${\bf s}$ as above, that the vector
$\frac{{\bf q}-{\bf r}} {|{\bf q}-{\bf r}|} + \frac{{\bf s}-{\bf r}}
{|{\bf s}-{\bf r}|} $ is normal to $M$ at ${\bf r}$. This suggests a method of constructing a
vector field ${\bf W}$ on $R^3$ that will be normal to the solution:
for each ${\bf r} \in R^3$ lying on a ray that enters the camera,
define

\begin{equation}
{\bf W}({\bf r})= \frac{{\bf q}({\bf r})-{\bf r}} {|{\bf q({\bf r})}-{\bf r}|} + \frac{
G({\bf q}({\bf r}))-{\bf r}} {|G({\bf q}({\bf r}))-{\bf r}|}
\end{equation}

\noindent
where ${\bf q}({\bf r})$ is the projection of {\bf r} to $I$ along the
ray. We refer to this construction as the {\bf vector field
method}. Thus our problem is solved if we find a surface whose
gradient is {\em parallel} to ${\bf W}$.

\section{Image Error and Projection Error}

A disadvantage of some of previous approaches  to the problem is that they do not directly address the error in the image, i.e., the goal should be to minimize the distortion error in the resulting image.

Given the notation that $T_M$ is the transformation induced from the
image plane to the object surface by the mirror $M$, then the goal is
to find a solution to the equations

\begin{equation}
T_M({\bf x}) = G({\bf x})
\end{equation}

\noindent
which is a system of partial differential equations that is generally
inconsistent (an example of such a system is equations \ref{eq:g1} and
\ref{eq:g2}). We then define the {\bf projection error} induced by a
mirror $M$ as

\begin{equation}
P_e(M) = \frac{1}{Area(U)}\int_U |T_M({\bf x})-G({\bf x})|^2 dA,
\end{equation}

\noindent
where $U$ is the domain in the image plane over which the surface $M$
is a graph. For example, in the approach to the blindspot problem
described below, the projection error of a mirror described as a graph
$x=f(y,z)$ over $[-1,1]\times[\-1, 1]$ is

\begin{equation}
\frac{1}{4}\int_{-1}^1 \int_{-1}^1 
\left( {\frac {\left (1 - {{\it f_y}}^{2} - {{\it f_z}}^{2}\right )\left (y+k
\right )}{2{\it f_y}}} + f(y,z) - \alpha y \right)^2 +
\left( -{\frac {\left (y+k\right ){\it f_z}}{{\it f_y}}}+z - \alpha z \right)^2
dydz,
\end{equation}

\noindent
where the prescribed transformation $G$ is $[x_0, y, z] \mapsto
[\alpha y, -k, \alpha z]$. (Here we are not taking the surface to be
at infinity.) For example, if $\alpha=1$ then the mirror $x=y$ makes
the projection error functional 0.

The projection error compares the image formed by projecting the
domain $U$ from the image plane to the object surface, whereas we are
interested in the error formed by the projection from the object plane
to the image plane via $M$. Thus we define the {\bf image error} as
the quantity

\begin{equation}
I_e(M) = \frac{1}{Area(V)}\int_V |G^{-1}({\bf y}) - T_M^{-1}({\bf y})|^2 dA,
\label{eq:ie}
\end{equation}

\noindent
where $V$ is the image of $U$ under $T_M$. In this form it is not
possible to directly minimize the image error because computing
$T_M^{-1}$ is intractable. While it is possible to compute the
projection for a generic $M$, it does not appear possible to compute
$T_M^{-1}$ because the structure of the bundle of rays from the object
plane to $M$ is unknown until $M$ is given. Nevertheless, performing a
change of variables on the integral (\ref{eq:ie}) by taking ${\bf
y}=T_M({\bf x})$ gives

\begin{equation}
I_e = \frac{1}{Area(V)}\int_V |G^{-1}({\bf y}) - T_M^{-1}({\bf y})|^2 dA = 
\frac{1}{Area(V)}\int_U  |G^{-1}(T_M({\bf x}))-{\bf x}|^2|\det(dT_M({\bf x}))|dA
\end{equation}

This functional is amenable to numerical minimization. One great
advantage of this functional is that any solution $M$ derived by {\em
any} one of the previous methods may be improved by minimizing $I_e$
with $M$ as an initial condition. 

\section{An application to design}

The main purpose of deriving an expression for the image error is so that it can be minimized over some appropriate family of surfaces and hence provide an answer to the prescribed projection problem.

In \cite{hicks01cvpr}, the authors consider the equations for a sideview mirror. It is shown that the projection $T_M$ induced by a surface $x=f(y,z)$ viewed orthographically along the x-axis (the image plane is $x=x_0$) to the plane $y=-k, k>0$ is  

\begin{equation}
[x_0, y, z] \rightarrow [g_1(y,z), -k, g_2(y, z)],
\end{equation}

\noindent
where

\begin{equation}
g_1 = {\frac {\left (1 - {{\it f_y}}^{2} - {{\it f_z}}^{2}\right )\left (y+k
\right )}{2{\it f_y}}} + f(y,z) 
\label{eq:g1}
\end{equation}

\begin{equation}
g_2 = -{\frac {\left (y+k\right ){\it f_z}}{{\it f_y}}}+z.
\label{eq:g2}
\end{equation}

The desired projection, $G$ is
\begin{equation} 
[x_0, y, z] \rightarrow [\alpha y, -k, \alpha z]
\end{equation}

Since $T_M$ and $G$ are known, we may minimize the image error over some class of surfaces. Ideally one would perform this minimization over a large space, such as polynomials, trigonometric functions, or spline functions. Here we answer the question "What is the best spherical sideview mirror ?".

We consider a spherical mirror which goes through the origin and with a center in the plane $z=0$, which has the general form

\begin{equation}
f(y,z) = -a + \sqrt{a^2-z^2-y^2+2yb}.
\end{equation} 

Thus the parameters that are free for minimization are  $a$ and $b$. For this problem, if the required field of view is 45 degrees (this determines $\alpha$ and assuming that the image plane is a unit square of side length 2, the optimal result, (using a gradient descendent algorithm) is 

\begin{equation}
a \sim 2.83
\end{equation}

\begin{equation}
b \sim 2.38.
\end{equation}

\renewcommand{\baselinestretch}{1.0}

\end{document}